\documentclass[conference]{IEEEtran}
\IEEEoverridecommandlockouts
\usepackage{cite}
\usepackage{amsmath,amssymb,amsfonts}
\usepackage{algorithmic}

\usepackage{amsthm}

\usepackage[ruled,vlined,linesnumbered]{algorithm2e}

\usepackage{graphicx}
\graphicspath{{./fig/}}
\usepackage{textcomp}
\usepackage{xcolor}
\usepackage{caption}
\usepackage{subcaption}

\def\BibTeX{{\rm B\kern-.05em{\sc i\kern-.025em b}\kern-.08em
    T\kern-.1667em\lower.7ex\hbox{E}\kern-.125emX}}
\begin{document}

\title{Network-Coding-based Forwarding for LoRaWAN  Gateways\\

}

\author{\IEEEauthorblockN{Louai Al-Awami}
\textit{Interdisciplinary Research Center for Intelligent Secure Systems} \\
\IEEEauthorblockA{\textit{Computer Engineering Department} \\
\textit{King Fahd University of Petroleum and Minerals}\\
Dhahran, Saudi Arabia \\
louai@kfupm.edu.sa}
\thanks{This project was funded by the Deanship of Research Oversight and Coordination (DROC) at King Fahd University of Petroleum and Minerals (KFUPM) through grant No. SR191011.}
}


\maketitle

\begin{abstract}
LoRaWAN is a promising IoT access technology that is growing in popularity. This study addresses the issue of duplicate packets forwarding by LoRaWAN gateways and proposes a novel forwarding scheme to eliminate forwarding duplicate packets by utilizing inter-flow network coding. The proposed scheme is distributed and requires no coordination between gateways. The proposed scheme is evaluated under different network and traffic conditions. The results show that substantial savings can be achieved in bandwidth leading to enhanced scalability of the network without increasing outgoing traffic.
\end{abstract}


\section{Introduction}
The Internet had an incomprehensible impact on our life but it seems as that was just the beginning. The next step in its evolution, or what is referred to as the Internet of Things (IoT), encompasses connecting traditionally unconnected things to the Internet paving the way for a smarter and a more efficient planet. IoT has already started to disrupt many sectors such as transportation, energy, health, and many more.

In order for IoT to succeed, a number of challenges must be overcome including efficiency in resource utilization, security, scalability, among others. Due to the sheer scale of the system - which may include billions of devices - any small gain in efficiency can quickly multiply. 

A key enabler of IoT is access technologies. Many access technologies have thus far been introduced to provide IoT devices with connectivity in short, medium, and long range applications. Long Range Wide Area Network (LoRaWAN) comes up as a strong contender in the Low-Power Wide Area Network (LPWAN) arena. LoRaWAN has shown a promising potential and received a huge attention from both the academic and the industrial communities.

\begin{figure}[ht]
	\centering
	\includegraphics[scale=0.6]{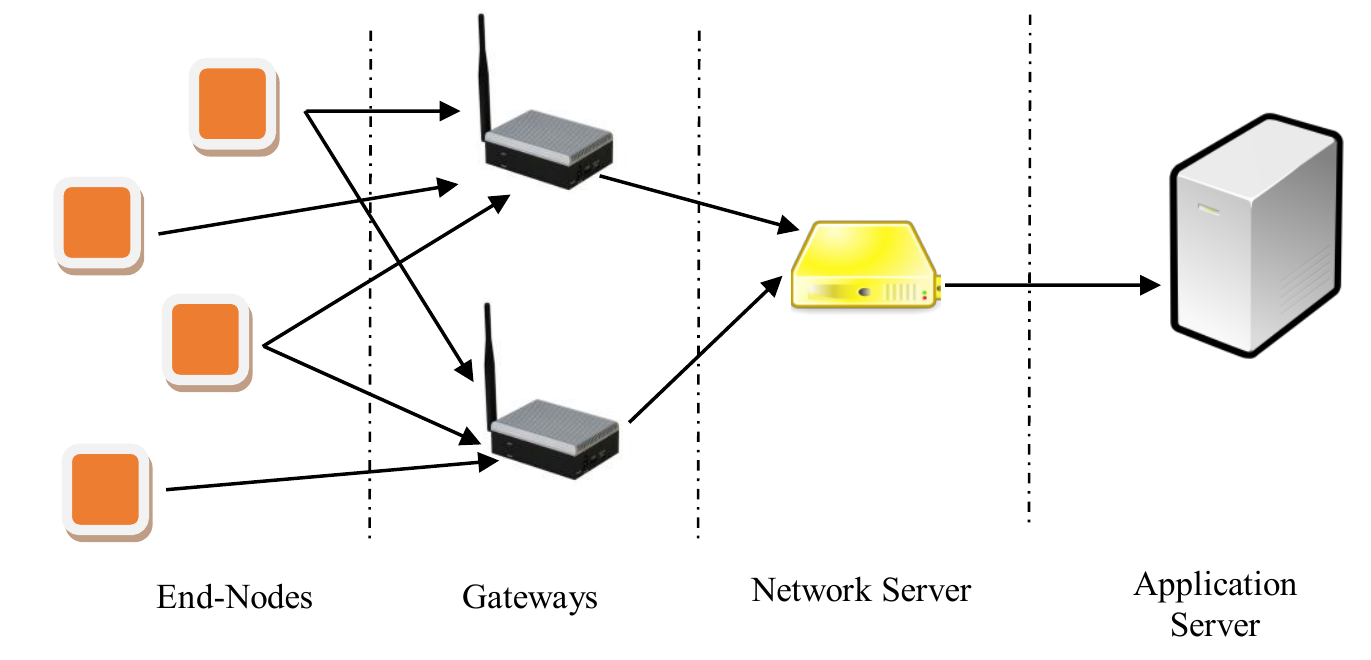}
	\caption{LoRaWAN Network Architecture.}
	\label{fig:lora-network}
\end{figure}

As seen in Fig. \ref{fig:lora-network}, LoRaWAN utilizes a star-of-stars topology where a set of sensor nodes communicate directly with a single or multiple gateways. The gateways, in turn, forward frames to a network server where most of the intelligence reside. LoRaWAN network servers are typically situated as a cloud service. Since gateways work independently, a frame received by multiple gateways results in duplicate packets traveling to the network server, where de-duplication takes place. 

As noted in \cite{Bor:2016} and \cite{VanDenAbeele2017}, LoRaWAN scalability can be greatly improved by increasing the number of gateways. For instance, in a network of $1600$ nodes, to achieve a packet delivery rate close to $90\%$, as many as $24$ gateways are needed \cite{Bor:2016}! However, as the number of nodes and gateways increases, duplicate packets can become a substantial portion of the data forwarded over the uplink to the network server. Furthermore, in case of remote deployment where back-haul bandwidth could be scarce, redundant traffic may result in an unnecessary strain on an already constrained system. The central motivation of this study is to introduce a network coding scheme that enables increasing the scalability of the network through increasing the number of gateways without increasing the amount of uplink data sent to the network server.

\begin{figure}
	\centering
	\begin{subfigure}[b]{0.4\textwidth}
		\centering
		\includegraphics[scale=0.55]{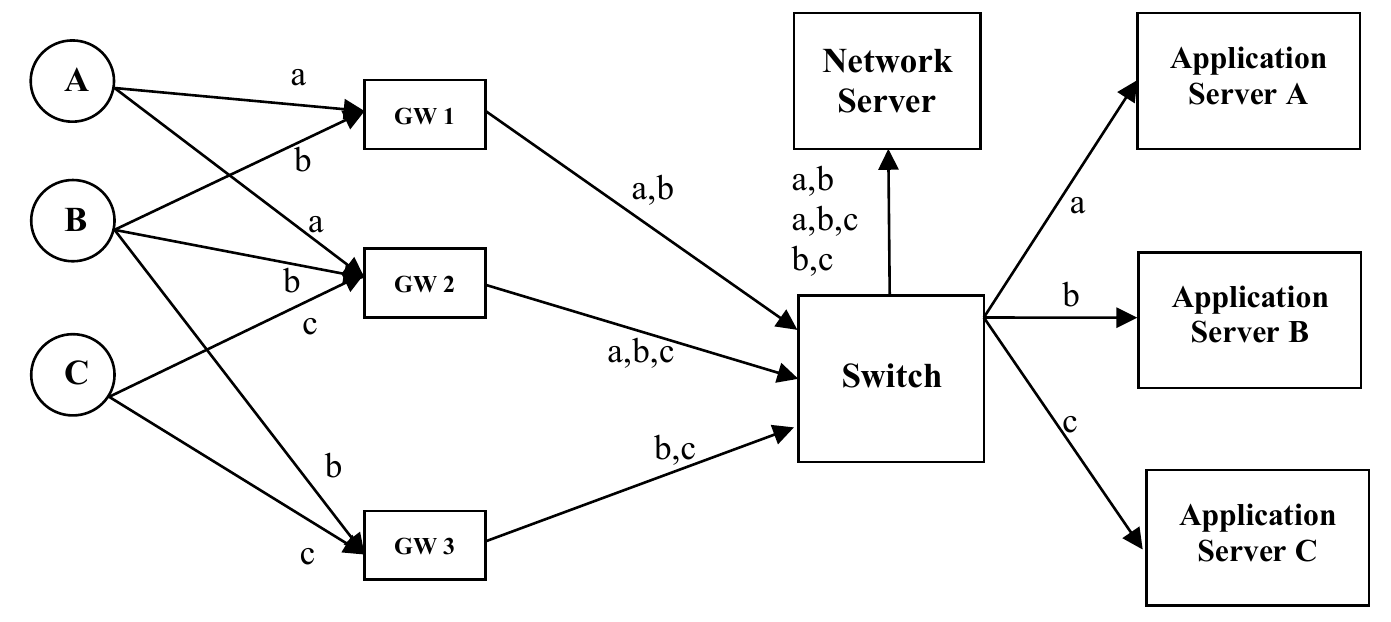}
		\caption{LoRaWAN uplink forwarding with no coding.}
		\label{fig:lorawan-no-coding}
	\end{subfigure}
	\hfill
	\begin{subfigure}[b]{0.5\textwidth}
		\centering
		\includegraphics[scale=0.55]{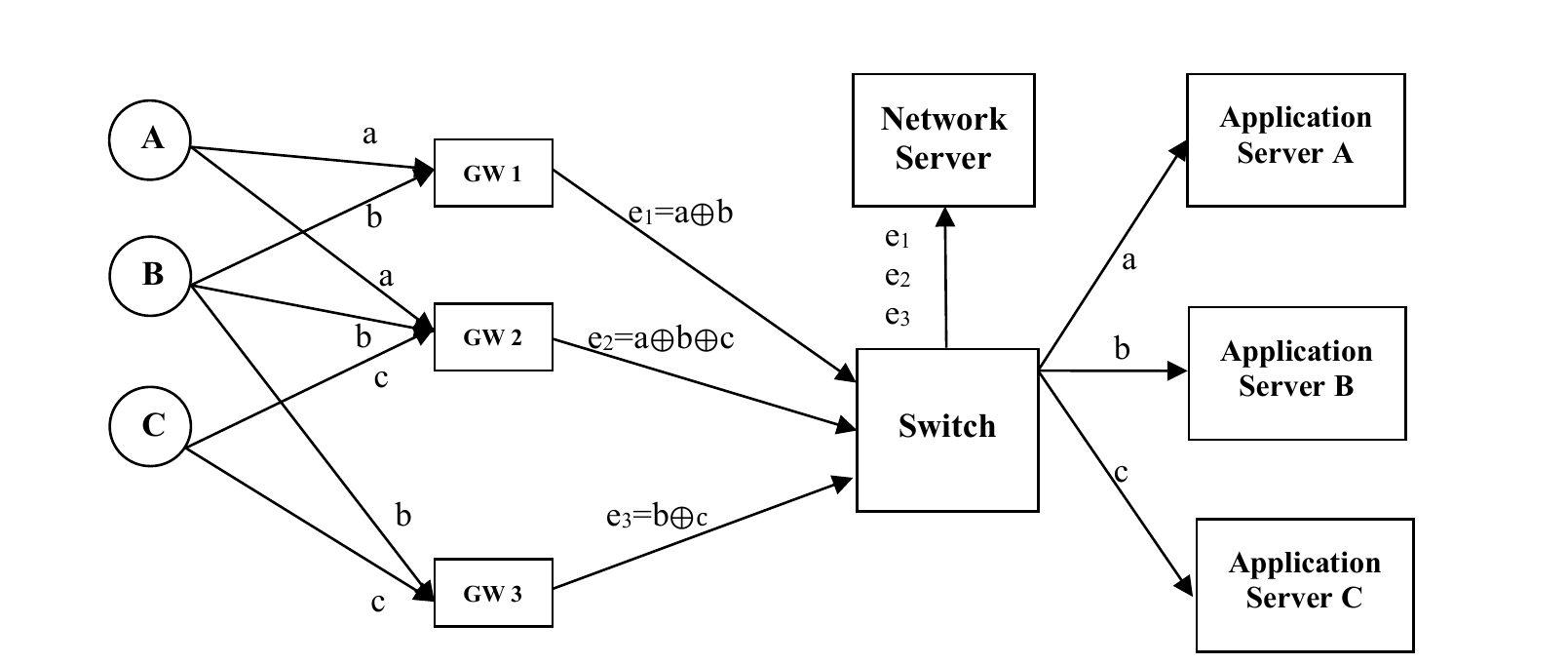}
		\caption{LoRaWAN uplink forwarding with coding.}
		\label{fig:lorawan-with-coding}
	\end{subfigure}
	\caption{LoRaWAN with and without network coding.}
	\label{fig:lora-toy-example}
\end{figure}

To motivate the benefits of using network coding, Fig. \ref{fig:lora-toy-example} shows a toy example. In this example, a network contains $3$ sensor nodes and $3$ gateways. Let $a$, $b$, and $c$ denote the packets sent by nodes $A$, $B$, and $C$, respectively. The lines between sensor nodes and gateways depicts the transmission range of each node. As dictated by the LoRaWAN standard, when the three nodes transmit their packets simultaneously, as in Fig. \ref{fig:lora-toy-example}-(a), a total of $7$ packets are forwarded to the network server because every gateway forwards all packets it receives. However, through the use of network coding, as in Fig. \ref{fig:lora-toy-example}-(b), the packets received can be combined resulting in reducing them to $3$ only, which translates to more than $50\%$ in bandwidth savings. Note that the size of an encoded packet, $e_i$, is equal to the size of one of the original packets. This observation forms the basis of the scheme proposed in this study.

Network coding  \cite{Ahlswede2000} has been shown to aid in bandwidth savings, increased throughput, reduced energy consumption, among other benefits. Many significant advancements have been made since in this field, most notably the introduction of Linear Network Coding \cite{Li2003} and Random Linear Network Coding (RLNC) \cite{chou2003practical}. The significant of the latter study is that it has shown that the benefits of network coding can be achieved using low complexity linear coding in a randomized setting which also enables distributed implementations. Network coding has received tremendous interest from the research community and has been used in many areas including wireless sensor networks \cite{Han2019}, distributed storage systems \cite{Al-Awami2016}, P2P networks \cite{AngelinJayanthi2014}, and more.

In this paper, a packet forwarding scheme is proposed utilizing random inter-flow network coding to reduce the traffic forwarded by LoRaWAN gateways by eliminating duplicate packets at the edge of the network. This enables using LoRaWAN in extreme low-bandwidth deployments without compromising the resolution of the transported data. Furthermore, the proposed scheme enables increasing the scalability of the network and reducing the traffic load on the cloud.

The contributions of this paper are as follows:
\begin{itemize}
	\item showing analytically how network coding can reduce the traffic of LoRaWAN gateways. An upper limit is derived on the possible reduction in traffic.
	\item proposing a distributed inter-flow network coding scheme to eliminate duplicate packets forwarded by LoRaWAN gateways; and evaluating and comparing its performance to that of the standard LoRaWAN forwarding mechanism through simulation.
\end{itemize}

The remainder of the paper is organized as follows. In Section \ref{section-bg}, a background of LoRaWAN and network coding is presented. Section \ref{section-survey} covers related work. The proposed scheme is presented in Section \ref{section-DFL} and the performance evaluation in Section \ref{section-performance}. Finally, Section \ref{section-conclusion} concludes the paper.

%

\section{Long Range Wide Area Network (LoRaWAN)}
\label{section-bg}

LoRaWAN \cite{LoRaWANAlliance2015} is a major proponent in the race to LPWAN IoT network access technology. LoRaWAN aims to provide long-range connectivity (beyond $10$ km) at low data rate ($< 50$ kbps) over the unlicensed ISM bands for IoT applications. The data link layer of LoRaWAN runs over the (Long Range) LoRa physical layer which was developed by the French company Cycleo and later acquired by Semtech in $2012$. The specifications of running LoRaWAN over Frequency Shift Keying (FSK) also exist. The design of LoRaWAN is optimized for range, scalability, cost, and battery life.

LoRa is a physical layer technology that operates in the ISM bands such as the $433$ MHz, $868$ MHz, and $915$ MHz bands. It is to be noted that operating bands differ by geographical region. LoRa uses a proprietary spread spectrum modulation technique derived from the Chirp Spread Spectrum (CSS) where a chirp is represented by a unique variation in frequency. The advantage of using CSS is that it eliminates the need for pseudo-random spread codes resulting in simpler receiver design. In addition, CSS is resilient to Doppler effect and multipath fading \cite{Semtech2015}. The scheme uses different spread factors which enables an Adaptive Data Rate (ADR) between $0.3$ and $50$ kbps.

LoRaWAN defines the upper layers of the technology including medium access control, networking, mobility management, security, etc. The LoRaWAN specifications are maintained by the LoRa Alliance \cite{LoRaWANAlliance2015}. The latest specifications as of this writing is version 1.1 \cite{LoRaWANAlliance2017}.

A LoRaWAN network is laid out as a star-of-stars topology. As shown in Fig. \ref{fig:lora-network}, end nodes communicate with gateways that forward messages to a network server which in turn routes the messages to the appropriate application server. Besides, the network server, with the help of a join server, is responsible for device activation (registration) and authentication. The link between the end nodes and the gateways uses the LoRaWAN protocol while the gateways, the network server, and the application server can communicate over any standard IP link.

A single transmission by a LoRaWAN end-node can be overheard, and therefore forwarded, by multiple gateways to increase the chance of successful reception. Thus, the network server is also responsible for de-duplication of redundant packets. This study addresses this specific aspect of the protocol as will be discussed shortly. LoRaWAN has mobility support as well. The network server can track the location of end devices through monitoring the gateway(s) the nodes use to forward their data. This same gateway is used to forward data and control from the application server to the corresponding end-device over the downlink direction.

LoRaWAN allows a bi-directional communication between devices and gateways. The standard defines three classes of devices with different requirements. Class A is for low-power devices and must be supported by all devices. An end device can initiate an uplink transmission at any time asynchronously. Each uplink transmission window is followed by two downlink windows in case the gateway needs to forward packets to the end device. In this class, a gateway can only send downlink data after the end device initiates an uplink communication. Thus an end device can enter a low power sleep mode for as long as the application requires with no need for regular wake ups. A Class B device sends periodic beacons to the network and opens downlink windows at regular intervals. The latency between transmissions can be set in software and can be up to $128$ s. This class is designed for application with minimum latency requirements. Finally, in Class C, an end device keeps the transceiver up all the time as to eliminate latency. This mode is for applications where latency is critical but energy is available. It is possible to switch between Class A and C if a large amount of data (such as Over-The-Air (OTA) firmware update) is being sent to an end device.

\section{Literature Survey}
\label{section-survey}
LoRaWAN has been seeing a growing interest from the research community in recent years. The authors in \cite{Haxhibeqiri2018}, provide an overview of the technology and survey the recent research in the area. In \cite{Sundaram2019}, the authors present a number of research problems and open issues in LoRaWAN.

Multiple efforts tried to improve LoRaWAN channel performance using coding. Most notably, \cite{Marcelis2017} develops an application layer coding scheme that encodes consecutive frames using erasure codes to improve recovering the transmitted data at the receiver. \cite{Sandell2019} analyzes further the proposal in \cite{Marcelis2017} using simulations. The paper concludes that due to limited duty cycling and the harsh nature of the channel, benefits of employing the proposed coding scheme at the physical LoRa link are limited. In \cite{COUTAUD2018}, a Reed-Solomon-based Forward Error Correction (FEC) is used to improve the Packet Delivery Ratio (PDR) in LoRaWAN networks. \cite{ratelessLora2020} investigates the use of rateless codes to improve LoRa communication range and reduce energy consumption. To overcome the restrictions of low-duty cycle of LoRaWAN, the authors in \cite{Borkotoky2019} introduced a redundancy scheme based on repetition coding to increase PDR in the absence of an acknowledgment mechanism. \cite{Montejo-Sanchez2019} proposes a packet-level coding scheme that utilizes erasure codes by encoding new messages with older messages to increase PDR. \cite{Bor:2016} and \cite{VanDenAbeele2017} evaluate the scalability of the protocol. The two papers raise doubts about the scalability of the medium access strategy used by LoRaWAN.

Most of the existing work focuses on improving the access layer through the incorporation of coding and redundancy. In contrast, this study seeks to improve the internal (backend) performance of LoRaWAN networks. To the best of the author's knowledge, this is the first study to address the issue of duplicate packets at the LoRaWAN gateways. While other related work focuses on packet delivery ratio on the physical LoRa links, this study aims to increase the efficiency and scalability of the system by eliminating duplicate packets and reducing the number of packets forwarded to network server.

\section{Proposed Scheme}
\label{section-DFL}
This section starts by introducing the notation used, system model, and assumptions. Then, it presents an analytical derivation of an upper limit on the possible gain achieved by network coding in LoRaWAN. Finally, it presents the proposed scheme.

\subsection{System Model}
Assume a LoRaWAN network to contain $n$ sensor nodes and $m$ gateways, where $n>m$. Let $U=\{u_1, u_2, ..., u_n\}$ and $V=\{v_1, v_2, ..., v_m\}$ be the set of sensor nodes and gateways in the network, respectively. Without loss of generality, time is assumed to be divided into generations where in each generation, a subset of sensor nodes transmit frames to the gateways. The subscripts $i$ and $j$ will be used to refer to sensor nodes and gateways, respectively. In each generation, each sensor node $u_i$ is assumed to transmit a maximum of one packet $x_i$ with a transmission probability $p_t$. $p_t$ is assumed to be equal for all nodes. Also, encoding takes place only between the set of packets, $X$, belonging to the same generation.

 To quantify the savings that can be achieved by network coding as opposed to pure forwarding, consider a single generation of packets. After successfully receiving packets from sensor nodes, a gateway can either forward them to the network server individually using \textit{pure forwarding}, or combine the packets using \textit{network coding} before forwarding them. The maximum number of packets is generated when all $n$ nodes transmit, and all $n$ packets are received by all $m$ gateways. Now, consider the following two cases:

\begin{itemize}
	\item \textbf{Case 1 - Pure Forwarding:} If no encoding is employed, all packets will be forwarded in raw form resulting in a total of $n\times m$ packets. 
	\item \textbf{Case 2 - Network Coding:} With network encoding, $n$ encoded packets are required to decode the $n$ native packets, therefore, the number of packets required to be forwarded to the network server is at least $n$ packets.
\end{itemize}

Let $B_{PF}$ and $B_{NC}$ be the bandwidth (in packets) required to forward all packets to the network server using pure forwarding and network coding, respectively. Then,
\[
B_{NC}=\left(\frac{n}{n \times m}\right) \times B_{PF}=\frac{B_{PF}}{m}
\]

This shows that if packets are combined by network coding at the gateways before being forwarded to the network server, then the bandwidth required to forward the packets can be reduced to $(\frac{1}{m})^{th} $ compared to that of pure forwarding. Equivalently, the savings in bandwidth resulting from using network coding compared to pure forwarding can be expressed as
\[
B_{PF}-B_{NC}=\left(1-\frac{1}{m}\right) \times B_{PF}
\]

Despite the glamorous results shown above, practical performance will vary depending on the number of nodes and gateways, traffic patterns, and network setup. Section \ref{section-performance} provides a detailed treatment of the performance of the system under various network settings and traffic patterns. 

\subsection{Network-Coding-based Forwarding (NCF)}

The proposed scheme is entitled Network-Coding-based Forwarding (NCF). It is comprised of three components: \textit{encoding vectors generation} at the network server, \textit{packet encoding} at the gateways, and finally \textit{packet decoding} at the network server.

The network server starts learning the nodes/gateways connectivity by observing the incoming packets. Connectivity can be inferred by noting which gateway forwards which packets. The network server then devises an encoding scheme which is forwarded to the gateways. The encoding scheme is represented by encoding vectors that are sent to each gateway. Accordingly, each gateway performs encoding on the packets it receives using the scheme received from the network server before sending them to the network server. Finally, the network server decodes the packets and recovers the original ones.

\subsubsection{Encoding Vectors Generation}
Given a LoRaWAN network with $n$ sensor nodes and $m$ gateways, the first step is for the network server to build a \textit{connectivity matrix} $(C)$ by observing the packets arriving from each gateway. The connectivity matrix is an $n \times m$ binary matrix which reflects the sensor nodes covered by each gateway. $C$ is later used to generate the set of \textit{encoding vectors} $G$. The general structure of $C$ is follows

\[
C = \begin{bmatrix} 
	c_{11} & c_{12} & \dots  & c_{1m}\\
	c_{21} & \ddots & & c_{2m}\\
	\vdots & &\ddots  &\vdots\\
	c_{n1} &        &  \dots& c_{nm} 
\end{bmatrix}
,\]

where

\[
c_{ij}=\left\{\begin{matrix}
	1 & \text{if gateway $v_j$ forwards packet from node $u_i$,}\\ 
	0, & \text{otherwise}. 
\end{matrix}\right.
\]
 
 Note that $C$ needs to be generated once only. However, it can be regenerated if nodes connectivity changes due to mobility, for instance. $C$ is then transformed and used to generate the set of encoding vectors $G$ as shown in Algorithm \ref{encoding-matrix-generation}.
 
 Let $C_{*j}$ denote the $j^{th}$ column vector of $C$. The \textit{weight of vector $C_{*j}$} (denoted as $W(C_{*j})$) refers to the number of non-zero entries in $C_{*j}$.  Note that $G_j$ in $G=\{G_1,G_2,..,G_m\}$, is a set of encoding vectors that is forwarded to gateway $v_j$ to be used to encode incoming packets.
 
  Algorithm \ref{encoding-matrix-generation} describes how the encoding vectors are generated from the connectivity matrix. For every node, a single entry is selected to remain as $1$ while the other entries are set to $0$ (lines $4-9$). This effectively instructs a specific gateway to include the packets coming from that node when encoding. The second loop (lines $11-16$) then computes the encoding coefficients and the number of encoded packets that need to be generated by each gateway per generation. The number of vectors generated by each gateway equals the weight of the column vector corresponding to $v_j$ in $C$.
 
\begin{algorithm}
	\SetKwInOut{Input}{Input}
	\SetKwInOut{Output}{Output}
	
	\Input{$C$: an $n \times m$ connectivity matrix. \newline
					$GF$: finite field.}
	\Output{$G$: a set of sets of $1 \times n$ encoding vectors.}
		$G \gets \{\varnothing\}$\;
		$F \gets \{\varnothing\}$\;
			\For{$j \gets 1$ to $m$}
		{  \For{$i \gets 1$ to $n$}{
				\If{$c_{ij} == 1$}{
						\If{$i \in F$}{
						$c_{ij} \gets 0$ \;
					}
					\Else{
						$F \gets F \cup \{i\}$\;
					}
					
				}	
			}
		$G_j \gets \{\varnothing\}$\;
		\For{$k \gets 1$ to $W(C_{*j})$}{  
			$g_{1\times n} \gets \vec{0}$\;
			\For{$i \gets 1$ to $n$}{
				\If{$c_{ij} == 1$}{
					$g_i \gets rand(GF)$ \;}
				}	
			
			$G_j \gets G_j \cup g_i$\;
		}
		$G \gets G \cup G_j$\;
			
		}
%
	return $G$\;
	\caption{Generation of Encoding Vectors}
	\label{encoding-matrix-generation}
\end{algorithm}

\subsubsection{Packet Encoding}
Now, gateways generate the encoded packets as described in Algorithm \ref{encoding-gateway}. Note that each round of the algorithm is applied to one generation of packets. It is assumed that at most one packet is received from any node per generation. The encoding involves summing up all packets received at the gateway over the finite field used after multiplying each packet with the random coefficient specified by the encoding vector. 

\begin{algorithm}
	\SetKwInOut{Input}{Input}
	\SetKwInOut{Output}{Output}
	
	\Input{$G_j$: a set of encoding vectors. \newline
		   $X$: a set of received packets in a generation.}
	\Output{$E_j$: a set of encoded packets.\newline
				 $\bar{G_j}$: a set of encoding vectors.}
	$E_j \gets \{\varnothing\}$;\\
	$\bar{G_j} \gets \{\varnothing\}$;\\
	$\bar{g}=$ first vector in $G_j$

	\ForEach{$x_i \in X$}{
	\If{$\bar{g}_i \neq 0$}{
		$g=$ fetch next vector in $G_j$;\\
		$e \gets \sum_{k=1}^{n}g_k . x_k $;\\
		 $E_j \gets E_j \cup \{e\}$\;
		 $\bar{G_j} \gets \bar{G_j} \cup g$\;	
}	
}
	return $E_j,\bar{G_j}$\;
	\caption{Encoding at Gateway $j$}
	\label{encoding-gateway}
\end{algorithm}

\subsubsection{Packet Decoding}

After receiving the encoded packets from gateways, the network server performs the decoding. Algorithm \ref{decoding-network-server} describes the steps used to decode the packets. Similar to the previous algorithm, decoding is applied to packets of the same generation. The decoding involves organizing the received vectors into a coefficient matrix $\bar{G}$ and column vector $E$. The original packets can then be found by multiplying $E$ and the inverse of $\bar{G}$. Note that the code in lines $3-5$ removes any redundant (linearly dependent) vectors.

\begin{algorithm}
	\SetKwInOut{Input}{Input}
	\SetKwInOut{Output}{Output}
	\Input{$E_1, E_2, ..., E_k$: a set of encoded packets.\newline
			$\bar{G}_1, \bar{G}_2, ..., \bar{G}_k$: a set of encoding vectors.}
	\Output{$X$: a vector of decoded packets.}
	$E= E_1 \cup E_2  \cup ... \cup E_k$;\\
	$\bar{G}=\bar{G}_1 \cup \bar{G}_2  \cup ... \cup \bar{G}_k$;\\
	
	\ForEach{$\bar{G_{*j}} \in \bar{G}$}{ 
		\If{$\bar{G_{*j}} == \vec{0}$}{
					remove $\bar{G}_{*j}$ from $\bar{G}$;\\
				}
			}
	$X = E . \bar{G}^{-1}$\;
	
	return $X$\;
	\caption{Decoding at Network Server}
	\label{decoding-network-server}
\end{algorithm}

%

\section{Performance Evaluation}
\label{section-performance}
To evaluate the proposed scheme, a custom simulator was developed. The simulator code and experimental data can all be found at \cite{NFC-simulator}. The goal of the simulation is to verify the theoretical gain under varying network configurations and traffic conditions. The simulator takes as input the number of gateways ($m$), the number of sensor nodes ($n$), the transmission probability ($p_t$), and the connectivity configuration. I have tested the scheme under two connectivity modes: RAND and EQUAL. Under RAND connectivity, the number of gateways reachable by every node is drawn uniformly at random from the set $\{1,2,...,m\}$. In contrast, EQUAL connectivity assumes all nodes can reach the same number of gateways. Under EQUAL connectivity, the number of gateways reachable by each node is denoted as the \textit{connectivity factor} ($w$). The simulator generates a random topology based on the input parameters. Using the random scenario generated, I tested both the proposed scheme, NCF, and the standard LoRaWAN protocol. The reported results were generated using $10,000$ runs with $95\%$ confidence level. The encoding in the simulator was under the mathematics of $GF=2^7$. In all scenarios, it is assumed that the number of gateways to be $5\%$ the number of sensor nodes, or $m=0.05 \times n$.

\begin{figure}[ht]
	\centering
	\includegraphics[scale=0.50]{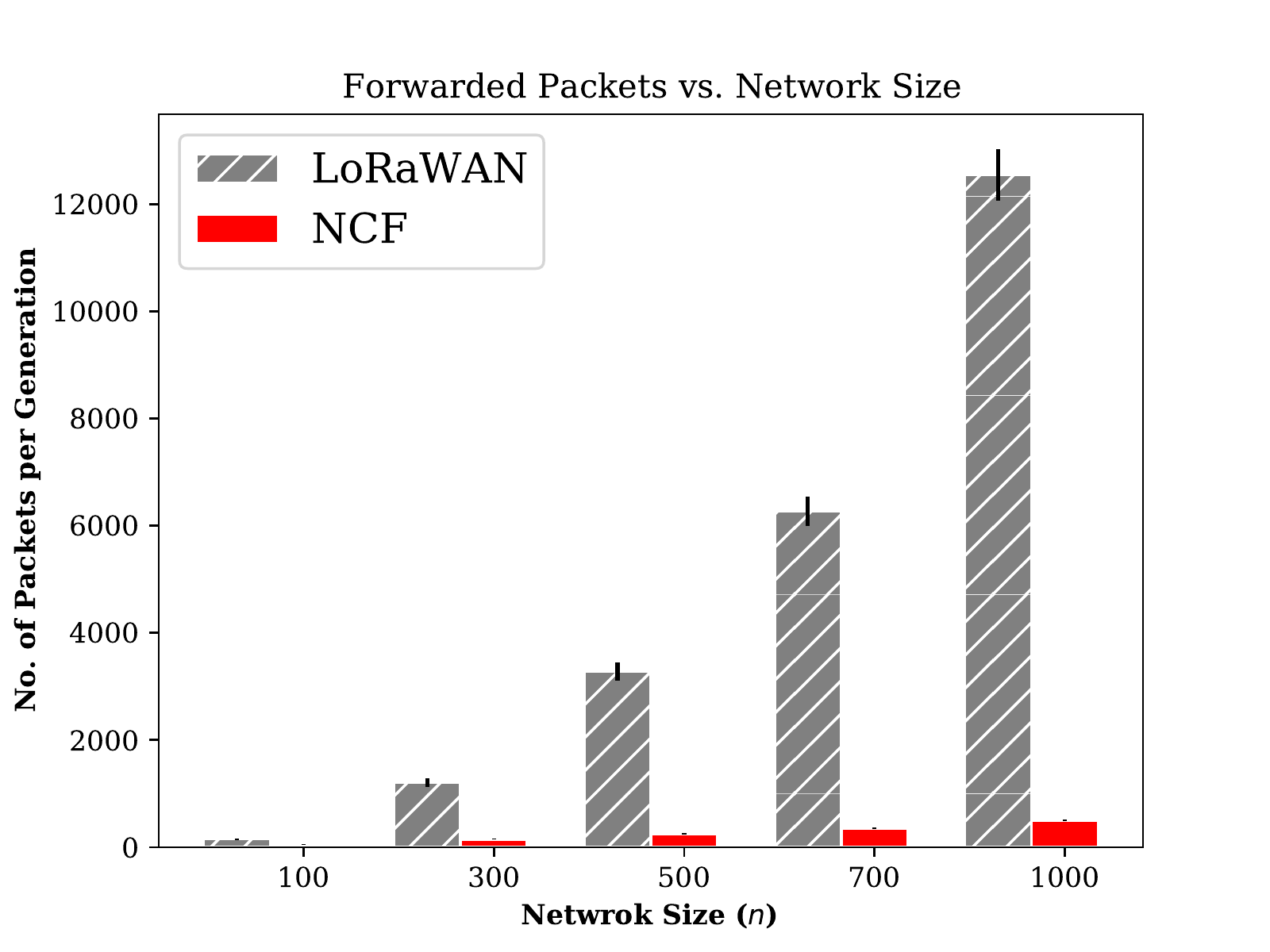}
	\caption{Packets forwarded versus network size ($p_t=0.5$, connectivity=RAND).}
	\label{fig:res-packets-vs-size}
\end{figure}

The performance metric used is the average packets per generation forwarded by the gateways to the network server. Fig. \ref{fig:res-packets-vs-size} shows the impact of the network size, on the number of packets forwarded by the gateways with and without network coding. This test was performed with a RAND connectivity and $p_t=0.5$. As seen in the figure, NCF can achieve substantial reduction in traffic compared to standard LoRaWAN. This confirms the analytical conclusion discussed earlier.

\begin{figure}[ht]
	\centering
	\includegraphics[scale=0.50]{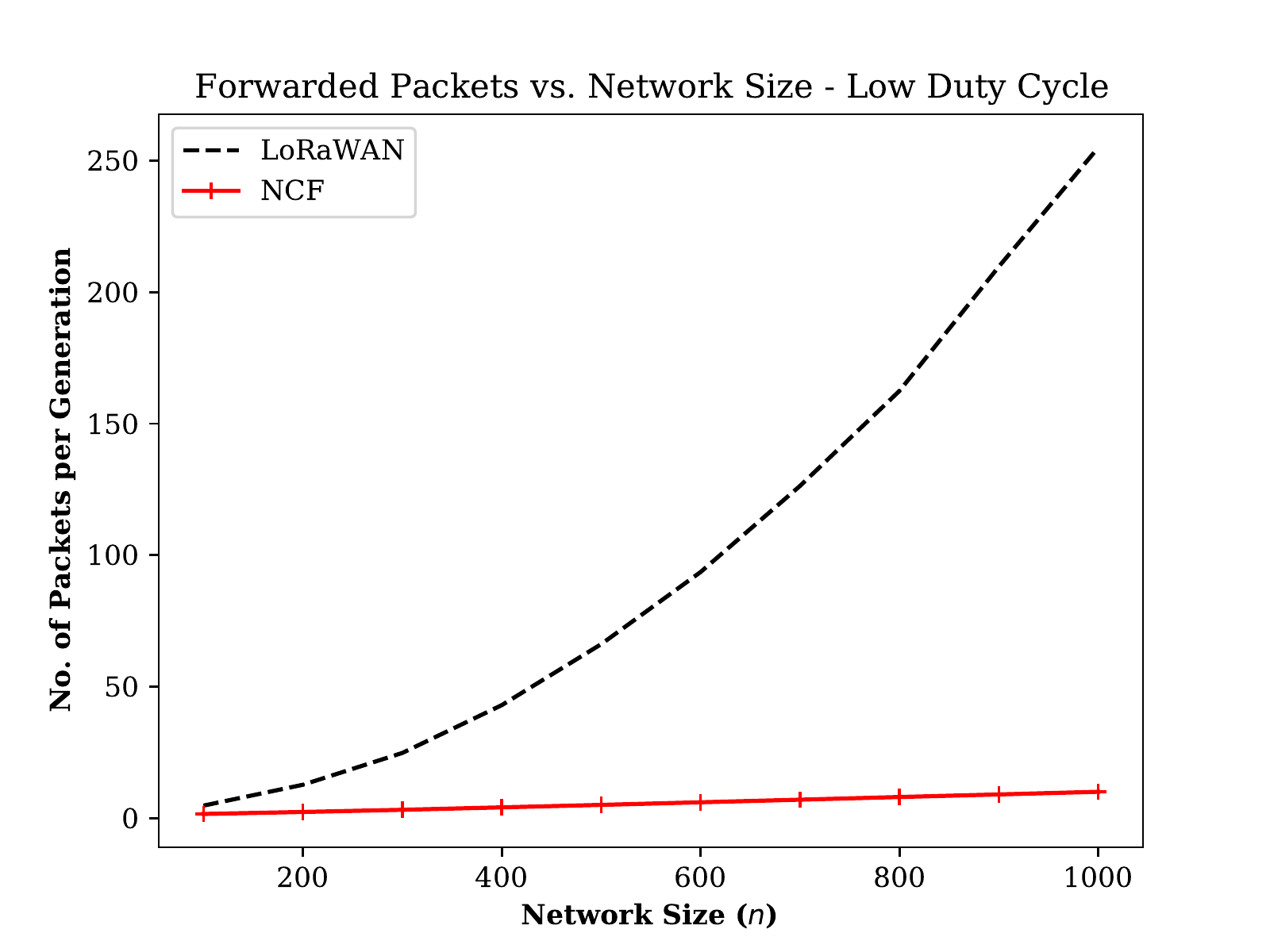}
	\caption{Packets forwarded versus network size under low duty cycle ($p_t=0.01$, connectivity=RAND).}
	\vspace*{-2mm}
	\label{fig:res-packets-vs-size_duty}
\end{figure}

LoRaWAN nodes often work under low-duty cycle to conserve power. In such a case, traffic may be generated at an extremely low rate. To show that NCF can still make a difference, Fig. \ref{fig:res-packets-vs-size_duty} shows the impact of network size on traffic under extremely low traffic ($p_t=0.01$). As shown, when the number of nodes increases, so does the traffic. However, NCF keeps the increase in check when compared to LoRaWAN.

\begin{figure}[ht]
	\centering
	\includegraphics[scale=0.50]{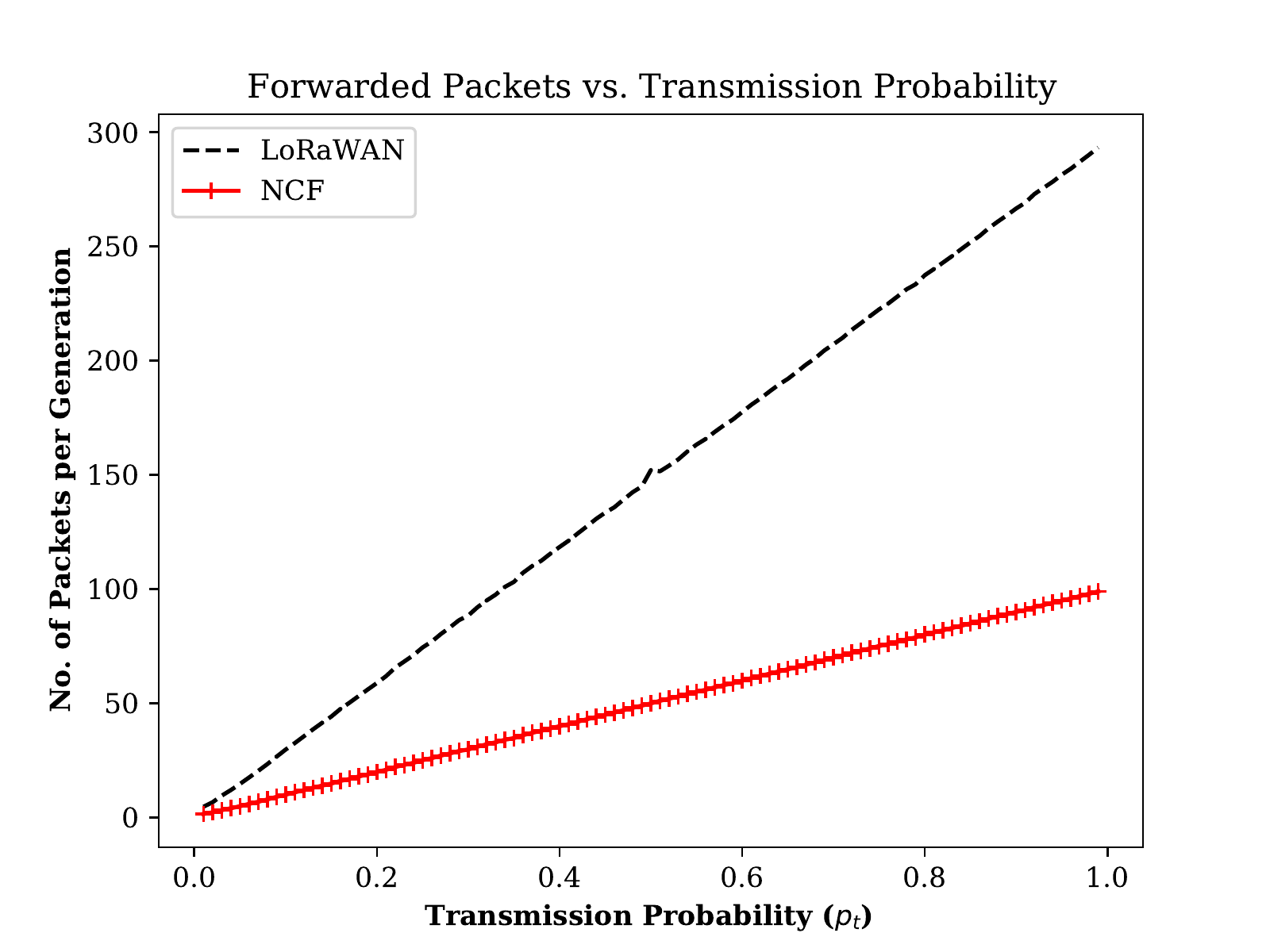}
	\caption{Packets forwarded versus transmission probability ($m=5$, $n=100$, connectivity=RAND).}
	\label{fig:res-packets-vs-tprop}
\end{figure}

It is known that network coding performance is limited by the arising of encoding opportunities. This was studied by examining the effect of rate of traffic on the performance of the two schemes. As seen in Fig. \ref{fig:res-packets-vs-tprop}, as the traffic increases (as a result of increasing $p_t$), the number of packets forwarded increases for both schemes. However, the rate of increase for LoRaWAN is significantly higher than that of NCF. For the sake of comparison, when $p_t=0.5$, NCF reduces the bandwidth requirements by $66\%$.

\begin{figure}[ht]
	\centering
	\includegraphics[scale=0.50]{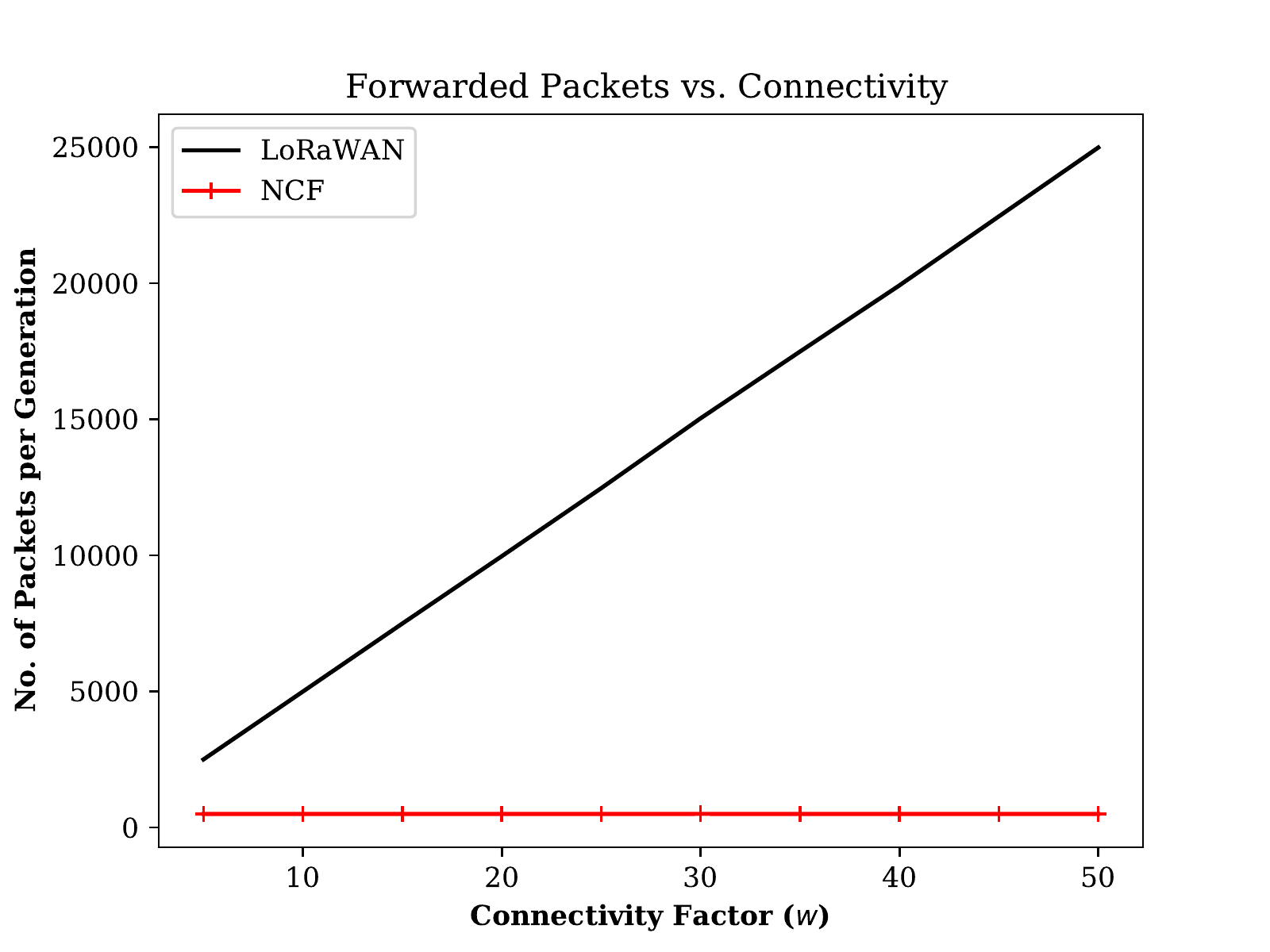}

	\caption{Packets forwarded versus connectivity ($n=1000$, $p_t=0.5$, connectivity=EQUAL.).}
	\label{fig:res-packets-vs-connectivity}
\end{figure}

Due to the low-power nature of LoRa nodes, multiple gateways are typically used to increase the chance of successfully receiving transmissions from sensor nodes. However, more gateways generate more duplicate packets which could limit the scalability of the network. Fig. \ref{fig:res-packets-vs-connectivity} shows how the impact of connectivity on the amount of traffic generated. What is surprising in the figure is that unlike standard LoRaWAN, increasing the connectivity factor in NCF does not affect the number of packets forwarded. This is logical since NCF removes duplicate packets while standard LoRaWAN does not. This demonstrates how NCF can indeed enhance the scalability of LoRaWAN through the utilization of network coding.

\section{Conclusion}
\label{section-conclusion}

In this paper, I have shown how the scalability of LoRaWAN can be enhanced by employing random inter-flow network coding at the gateways. I have also examined the achievable gain and the impact of the traffic rate and node connectivity on the forwarded traffic. The results confirm that as high as $66\%$ gain is possible under moderate settings. Higher gains are also possible. As a future extension, it would be interesting to examine the encoding scheme under varying channel and topology conditions.

%


\bibliographystyle{IEEEtran}
\bibliography{IEEEabrv,paper}

\end{document}